\documentclass[aps,prl,twocolumn,showpacs,preprintnumbers,amsmath,amssymb,superscriptaddress,floatfix]{revtex4}

\usepackage{graphicx}% Include figure files
\usepackage{dcolumn}% Align table columns on decimal point
\usepackage{bm}% bold math

\begin{document}

\title{Enhanced dielectric response by disordered nanoscale/mesoscopic
insulators}

\author{Shigeki Onoda}
\email{sonoda@appi.t.u-tokyo.ac.jp}
\affiliation{Spin Superstructure Project, ERATO, Japan Science and
Technology Agency, \\
c/o Department of Applied Physics, University of Tokyo, 7-3-1,
Hongo, Tokyo 113-8656, Japan}

\author{Chyh-Hong Chern}
\affiliation{Spin Superstructure Project, ERATO, Japan Science and
Technology Agency, \\
c/o Department of Applied Physics, University of Tokyo, 7-3-1,
Hongo, Tokyo 113-8656, Japan}

\author{Shuichi Murakami}
\affiliation{CREST, Department of Applied Physics, University of
Tokyo, 7-3-1, Hongo, Tokyo 113-8656, Japan}

\author{Yasushi Ogimoto}
\affiliation{Correlated Electron Research Center, National
Institute of Advanced Industrial Science and Technology, 1-1-1,
Higashi, Tsukuba, Ibaraki 305-8562}

\author{Naoto Nagaosa}
\affiliation{CREST, Department of Applied Physics, University of
Tokyo, 7-3-1, Hongo, Tokyo 113-8656, Japan}
\affiliation{Correlated Electron Research Center, National
Institute of Advanced Industrial Science and Technology, 1-1-1,
Higashi, Tsukuba, Ibaraki 305-8562}

\date{\today}

\begin{abstract}
Enhancement of the dielectric response of insulators by disorder
is theoretically proposed, where the quantum interference of 
electronic waves through the nanoscale/mesoscopic system 
and its change due to external perturbations control the polarization.
In the disordered case with all the states being localized, 
the resonant tunneling, which is topologically protected, 
plays a crucial role, and enhances the dielectric response by a factor 30$\sim$40 compared with the pure case. Realization of this idea with accessible materials/structures is also discussed.
\end{abstract}

\pacs{
03.65.Vf, %Phases: geometric; dynamic or topological
72.15.Rn, %Localization effects (Anderson or weak localization)
73.63.Nm, %Quantum wires (73.63.-b Electronic transport in nanoscale materials and structures (see also 73.23.-b Electronic transport in mesoscopic systems))
77.22.Ej, %Polarization and depolarization (77.22.-d, %Dielectric properties of solids and liquids)
85.35.Be %Quantum well devices (quantum dots, quantum wires, etc.) (85.35.-p Nanoelectronic devices)
}
\maketitle 

Dielectric property of insulators has been one of the most fundamental issues in condensed matter physics. Recently, it has been utilized as the ferroelectric random access memory (FeRAM)~\cite{Scott}, where thin films of ferroelectrics are integrated onto semiconductor Si wafers. The current FeRAM is comprised of a metal-insulator-metal structure illustrated in Fig.~1~(a). Electric polarization is reversed by applying an anti-parallel voltage pulse, trasferring from one lead to the other the electric charge determined by $2P_r$ with $P_r$ being the remnant polarization. By contrast, applying a parallel voltage pulse does not yield a polarization change or a charge transfer. This polarization current is the operation principle of the current FeRAM. Especially thin ferroelectric films of the nanoscale thickness are used, where the quantum effect becomes more important. Therefore, a quantum theory of nanoscale/mesoscopic capacitance is called for. For applications, it is also often required to achieve (i) a magnitude of the polarization larger than 10 $\mu$C/cm$^2$, (ii) a smaller leak current than 0.1-1 $\mu$A/cm$^2$, and (iii) a dielectric constant larger than 300~\cite{Scott}.
\begin{figure}[ht]
  \begin{center}\leavevmode
    \includegraphics[width=8.4cm]{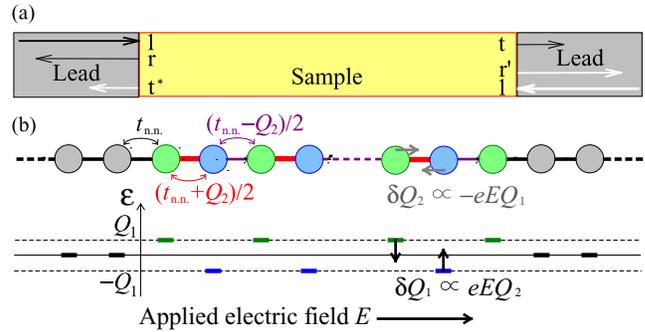}
  \end{center}
 \caption{\label{fig:model}(Color online) Ionic dimer system sandwitched by
the leads. (a) Transmission and reflection coefficients,  $t$ ($t'$) and $r$ ($r'$) at the left (right) edge of the sample. (b) Ionic level alternation ($Q_1$) and bond dimerization ($Q_2$), and effects of electric field $E$ on the changes $\delta\vec{Q}$ on $\vec{Q}$.}
\end{figure}

Generally, the dielectric response is inversely proportional to the energy gap $E_G$~\cite{Thouless83,OnodaMurakamiNagaosa04}. Here, $E_G$ is the charge excitation gap of the system. Therefore, a small charge gap is required to enhance the polarization and the dielectric response. One of the most advantageous ways of achieving this is to introduce a disorder potential~\cite{NiuThouless84} as realized in solid-solution systems, since neither dc transport nor dissipation by leak current occurs due to the localization effect~\cite{Anderson58,LeeRMP}. In fact, the localization of the wavefunction disconnects a coherent charge displacement and undesirably leads to a charge accumulation. This difficulty is overcome by taking nanoscale/mesoscopic capacitors with the thickness of the order of the localization length.

In this Letter, we develop a quantum theory of small-size capacitance, 
where the coherence of the electronic waves through the sample matters. 
Based on this, we propose a disorder-induced quantum-mechanical enhancement 
of the dielectric response in nanoscale/mesoscopic ferroelectrics, 
which provides a novel efficient principle of the FeRAM 
rapidly working with low dissipation. 
It also reveals topological properties of the ferroelectrics: 
in the presence of the disorder, a perfect transmission or equivalently 
a vortex of the reflection coefficients exists 
as the resonant tunneling~\cite{Azbel1983} in the space of 
the ionic gap and the dimerization.
Around this vortex center, the most efficient charge transfer 
can be achieved by controling the phase of the reflection coefficient,
as has been discussed in the context of quantum dots~\cite{resonant}.

We consider a one-dimensional (single-channel) insulating electronic system 
sandwiched by leads (electrodes) as shown in Fig.~1~(a). 
The extension to higher dimensional (multi-channel) cases will be
discussed later. We adopt the Landauer-B{\"u}ttiker
formalism~\cite{LandauerButtiker,Aharony}
where the sample is regarded as a {\it scatterer} characterized by
the scattering $S$-matrix 
$ S = \left(\begin{array}{cc}
 r & t \\
 t^* & r'
 \end{array}\right).
$ 
The transmittance through the sample and the reflectance at the
both ends are $T = |t|^2$ and $R = |r|^2 = |r'|^2=1-T$,
respectively. Then, we employ the Brouwer's
formula~\cite{Brouwer98}, which has been successfully applied to
the charge pumping in metallic quantum dot systems~\cite{Marcus99}: 
the charge $\Delta q$ pumped from the left lead to the right 
during an adiabatic change of parameters $\vec{Q}$ along a path $C$ 
is given by~\cite{Brouwer98}
\begin{equation}
 \Delta q = e\int_C\!\frac{d \vec{Q}}{2\pi} \cdot {\rm Im}
 \left(r^*\vec{\nabla}_Q r + t\vec{\nabla}_Q t^*\right).
 \label{eq:q}
\end{equation}
Therefore, even in the perfectly reflecting case $t=0$,
charge pumping or transfer occurs from one lead to the other by controlling the phase $\varphi$ of $r=|r|e^{i\varphi}$~\cite{resonant,Avron2004}.

We take the Hamiltonian schematically shown in Fig.~\ref{fig:model}~(b): $H = H_s + H_{\mathrm{lead}}$, where
\begin{eqnarray}
 H_s &=& -\sum_{i=1}^{L-1}\frac{t_{\mathrm{n.n.}}-(-)^iQ_2}{2}
(c^\dagger_{i+1}c_i+h.c.)
\nonumber\\
&&\makebox[1cm]{} 
+\sum_{i=1}^L ((-)^{i}Q_1+v_i) c^\dagger_ic_i,
 \label{eq:H_s}
\end{eqnarray}
for the sample and $H_{\mathrm{lead}}$ for the leads and interfaces 
is obtained by putting $Q_1=Q_2=v_i=0$ in $H_s$.
Here, $c_i$ and $c_i^\dagger$ are the annihilation and creation
operators of the electron at the site $i$. 
$t_{\mathrm{n.n.}}$ is the transfer integral. 
$Q_1$ and $Q_2$ represent the alternations of the local ionic level 
and the bond dimerization, respectively. 
$v_i$ denotes a local random potential at the site $i=1,\cdots,L$ 
with $L$ being the number of sites in the sample. 
The spin degree of freedom is omitted for simplicity. 
$H_s$ describes the essence of the electric polarization, i.e., 
the relative displacement of the two species of atoms, 
and describes the ferroelectricity in BaTiO$_3$~\cite{Ishihara93}
and quasi-one dimensional ferroelectrics 
such as organic charge transfer compounds
TTF-CA~\cite{ni} and (TMTTF)$_2$X~\cite{tmttf}.
Experimentally, one can change the parameters $\vec{Q}$ by applying
the electric field $E$ along the polarization direction and the
pressure $p$. The electric field $E$ linearly changes the angle 
$\theta=\arctan(Q_2/Q_1)$:  
it can be viewed that electrons at high and low deinsity sites shift in relatively opposite directions, as shown in gray arrows of Fig.~\ref{fig:model}~(b), changing the dimerization $Q_2$ by $\delta Q_2\propto eEQ_1$. Simultaneously, within each dimer, a level difference $Q_1$ changes by $\delta Q_1\propto-eEQ_2$ as illustrated by black arrows in Fig.~\ref{fig:model}~(b).
Applying the pressure reduces the magnitude of the gap $\sqrt{Q_1^2 + Q_2^2}$ through the increase of hybridization.

\begin{figure}[tbp!]
  \begin{center}\leavevmode
    \includegraphics[width=8.0cm]{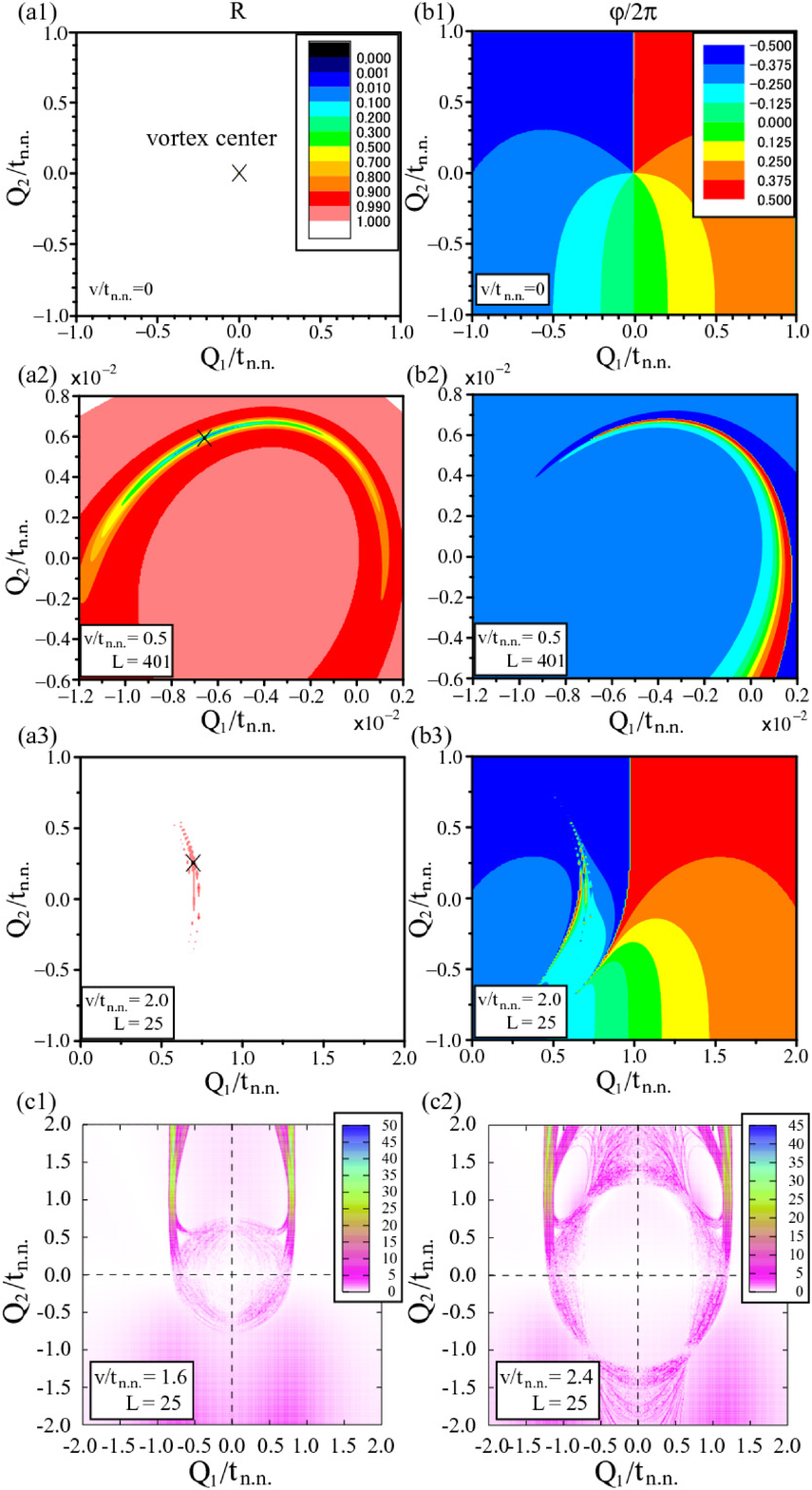}
    \includegraphics[width=8.0cm]{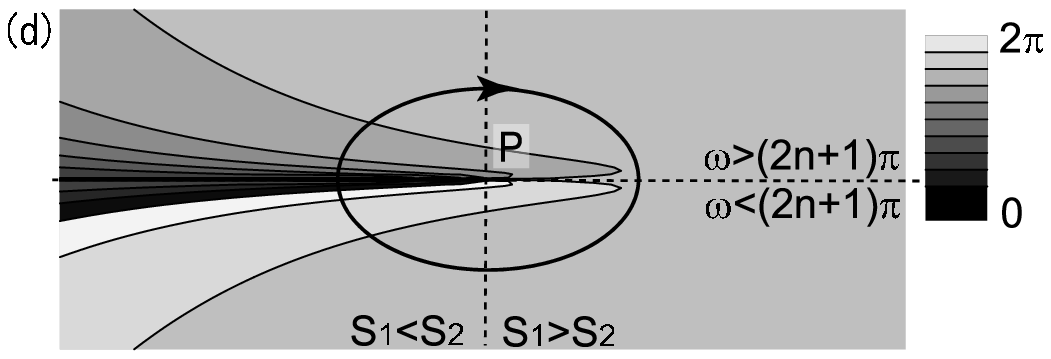}
  \end{center}
 \caption{\label{fig:r-phase} (Color) Reflectance $R=|r|^2$, and the phase
$\varphi={\rm arg}\ r$. (a1), (b1): clean bulk system. (a2), (b2):
disordered system of finite size $L=401$ with the random potential
of the strength $v/t_{n.n}=0.5$. (a3), (b3): disordered system with
$L=25$ and a stronger randomness $v/t_{n.n.}=2.0$ (see the text). The
color code given in (a1)/(b1) also applies to (a2,3)/(b2,3). In
the white region in (a1,2,3), $R$ is the unity within the accuracy
of $10^{-4}$. (c1,2): relative dielectric response in the presence of the disorder compared with the pure case. The disorder strength is $v/t_{n.n.}=1.6$ for (c1) and $2.4$ for (c2). Averages are taken over $10^2$ random disorder configurations. Inside white bands, there occurs a gradual sign change in the dielectric response. (d): Schematic contour mapping of the phase $\varphi$ of $r$.}
\end{figure}

Let us start with the pure case ($v_i=0$). Two uppermost panels of
Fig.~\ref{fig:r-phase} show the reflectivity $R=|r|^2$ for
$L=10001$ and the phase $\varphi$ of $r$ for $L\to\infty$ without
randomness. Here the phase winds by $2 \pi$ around
$\vec{Q}=(0,0)$, forming a ``vortex'' at which $r=0$ and $|t|=1$.
(To be precise, the transmittance $T$ behaves as $e^{-
L/\xi_{0}}$ with $\xi_0 = t_{\mathrm{n.n.}}/E_{G0}$. Therefore the
size of the region with a large $T$ is of the order of
$t_{\mathrm{n.n.}}/L$ for a large $L$.) To understand this result,
it is helpful to consider the bulk system. Without the random
potential $v_i$, the Hamiltonian $H_s$ under the periodic boundary
condition yields a gap 
$E_{G0} = 2 \sqrt{ Q_1^2 + {\rm Min}\{t_{\mathrm{n.n.}}^2,Q_2^2\}}$, 
which closes at $\vec{Q}={\vec 0}$~\cite{OnodaMurakamiNagaosa04}. 
Namely the vortex corresponds to the gapless case, 
where the extended state at the Fermi energy carries the charge 
and causes the perfect transmittance. 
This perfect-transmission point $\vec{Q}=0$ governs 
topological properties of the system in the whole $\vec{Q}$ plane 
in a non-local way. Along a cycle around $\vec{Q}={\vec 0}$, 
the charge $e$ is pumped and the polarization changes 
by $\pm 2ea$ ($a$: lattice constant) according 
to the Brouwer's formula Eq.~(\ref{eq:q}), since $t$ vanishes. 
This quantized charge pumping~\cite{Thouless83} 
is completely consistent with the results obtained 
in the periodic system~\cite{OnodaMurakamiNagaosa04,MurakamiOnodaNagaosa06} 
by using the Berry-curvature formulation of 
the electric polarization~\cite{rmp_Resta}.

Let us introduce the disorder. In the ${\vec Q}$ plane, 
the region of $|{\vec Q}| \lesssim v$ becomes gapless. 
The Anderson localization seriously affects
the transport properties~\cite{Anderson58,LeeRMP}. In one dimension,
the effect is pronounced and all the states are
localized~\cite{Wegner76,weaklocalization79,MacKinnonKramer81}. On
the other hand, the total vorticity is an integer
topological number robust under a continuous change of
parameters: the phase of $r$ ($r'$) should wind by $(-)2 \pi$ for a large
cycle far from $\vec{Q}=0$, well within the gapped region~\cite{NiuThouless84}.
There should be at least one vortex ($r=0$) in the
$\vec{Q}$ plane. Therefore, perfect transmittance $|t|=1$ occurs even
though all the states are strongly localized.

Figures \ref{fig:r-phase} (a2) and (b2) show our numerical results 
of $R$ and $\varphi$ for $L=401$ 
with a uniform random distribution $v_i\in[-v/2,v/2]$ for
the disorder strength $v=0.5t_{\mathrm{n.n.}}$. The location of the
perfect transmittance $|t|=1$ shifts in the $\vec{Q}$ space.
Besides, the shape of the region of relatively high transmittance
changes from isotropic to highly anisotropic.
In this anisotropic ``wing", $\varphi$ changes rapidly. 
The width $Q_W$ of the anisotropic wing decays exponentially 
as $\exp(-L/\xi)$ with incresing $L$. 
Here, the localization length $\xi$ is obtained from a numerical 
calculation of the inverse participation ratio 
as $\xi = (\sum_i^L D_i)^2/\sum_i D_i^2$, 
where $D_i$ is the local density of states at the site $i$. 
We obtain $\xi =76.4$ in the case of Figs.~\ref{fig:r-phase} (a2) and (b2).
In Figs.~\ref{fig:r-phase} (a3) and (b3), 
we further increase the strength of the random potential
by taking $v_i= s_i(v+\delta v_i)/2$ with $v/t_{\mathrm{n.n.}}=2.0$
for a smaller sample size $L=25$, 
where $s_i=\pm1$ denotes a random sign and 
$\delta v_i/v\in[-0.05,0.05]$ a uniform random distribution.
This type of randomness mimics the effects of substitution in alloys.
In this case, the region in the $\vec{Q}$ plane where the gap collapses and
vortices distribute expands. The localization length
$\xi \sim3$ or 4 is much shorter, and the transmittance $T$
is typically $\sim10^{-9}$, which is practically negligible,
except at the vortex core.

Properties around the perfect transmittance can be explained in
terms of the resonance tunneling \cite{Azbel1983,resonant}. When the Fermi
energy coincides with one of the eigenenergies of localized
states, the conductance is resonantly enhanced due to tunneling
from the usual exponentially suppressed value for $\xi\ll L$. 
For an effective potential $V(x)$ having two peaks with their heights 
characterized by the dimensionless parameters $S_i$ ($i=1,2$)~\cite{Azbel1983},
the reflection coefficient at the left end reads
$r=e^{i\phi} \left(\tanh S_{1}+\tanh S_{2}e^{i\omega}\right)/
\left(1+\tanh S_{1}\tanh S_{2}e^{i\omega}\right)$, 
with a real constant $\phi$, $\omega= 2\int_{x'}^{x''}k(x)dx$ 
in the semiclassical approximation~\cite{Azbel1983}, 
and an effective wavenumber $k(x)\sim \sqrt{E_F-V(x)}$. 
($S_1$ and $S_2$ are regarded as effective parameters 
when the real potential is much more complicated as in our case. 
They may be directly relevant to double-dot systems~\cite{double-dot}.) 
Then, a perfect transmission $|t|=1$, or equivalently a vortex or $r$, occurs 
when $S_{1}=S_{2}$ and $\omega=(2n+1)\pi$ ($n$:integer). 
The latter is the Bohr quantization condition that the
energy of the localized eigenstate is equal to $E_F$~\cite{Azbel1983}. 
In a cycle going around the vortex center in the direction to the arrows, 
the phase of $r$ winds by $2\pi$, as shown in Fig.~\ref{fig:r-phase}(d), 
while $r'$ yields the opposite phase winding. 
Since the transmittance is negligibly small except at the vortex center, 
according to the Brouwer's formula Eq.~(\ref{eq:q}),
this adiabatic cycle pumps nearly one unit charge from
the left lead to the right, like a bicycle pump~\cite{Avron2004}.
This nearly quantized charge pumping was discussed 
in the context of quantum dots~\cite{resonant}. 

When $\xi\ll L$, $S_1$ and $S_2$ are of the order of $L/\xi\gg1$, 
and the phases of $r$ and $r'$ abruptly change 
by $2\pi$ within the range of $|\omega - (2n+1)\pi| \sim e^{- L/\xi}$. 
Along the line $\omega=(2n+1) \pi$, we obtain $|r|= \tanh |S_1-S_2|$, 
which appreciably deviates from zero only when $|S_1-S_2| \gtrsim 1$, 
i.e., in a small region of the order of $\Delta/t_{\mathrm{n.n.}} \sim \xi/L$. 
These explain our numerical results. In such case,
because the tunneling rate is exponentially small as $Q_W \sim
t_{\mathrm{n.n.}} e^{-L/\xi} $, an exponentially long time
$\tau \sim (\hbar/t_{\mathrm{n.n.}}) e^{L/\xi}$ is required to
adiabatically pump charge, refleecting the uncertainty between
energy and time: in order to define the parameters $\vec{Q}$
with the higher accuracy than $Q_W$, 
we need to change $\vec{Q}$ in a longer time than $\hbar/Q_W$. 
If the parameters are changed in a high frequency $\omega>Q_W/\hbar$, 
the pumped charge obtains exponentially decaying factor of
$E_G/\omega$ as a nonadiabatic correction~\cite{ShihNiu94}.
This behavior resembles the relaxor ferroelectrics~\cite{Cross87,science_relaxor,Horiuchi00}.

To attain the enhanced dielectric response, 
the system should be located close to the ``wing'' in the $\vec{Q}$ space.
We consider a thin film consisting of many channels,
each of which is described by the present disordered model 
but with a different profile of random potentials. 
We can design the pattern of the phase $\varphi$ 
in the $\vec{Q}$ plane by the disorder. 
For the alloy model with the random on-site potentials,
vortices are mostly located around $Q_1=\pm v/2$
with the ``wing'' almost parallel to the $Q_2$ axis.
Therefore, we can enhance the dielectric response 
by tuning the disorder strength 
so that the realized $\vec{Q}$ point is located in the ``wing'' region.
In particular, when $T$ is negligibly small, 
the enhancement factor is given by $(\partial\varphi/\partial\theta)_{\rm disorder}/(\partial\varphi/\partial\theta)_{\rm pure}$ from Eq.~(\ref{eq:q}), since the electric field gives a linaer change in $\theta=\arctan(Q_1/Q_2)$.
Figures~\ref{fig:r-phase}(c1) and (c2) represent the $\vec{Q}$ dependence 
of the enhancement factor in the case of $v/t_{n.n.}=1.6$ and $2.4$ 
for $L=25$, respectively, with the number of channels being $10^2$.
The results reveal that around $Q_1\sim\pm v/2$, the dielectric response 
is largely enhanced by a factor $30\sim40$ compared with the pure case.
Even for the thin film with a square shape of a linear dimension 
larger than $50 \mbox{\AA}$, which corresponds to $N=10^2$, 
the disorder-driven enhancement of the charge tranfer rate remains robust.
This allows for a reduction of the applied electric field necessary
for manipulating the polarization.
If we require the response time $\tau\sim e^{L/\xi}/t_{\mathrm{n.n.}}$
of the order of $10^{-9}\ {\rm s}$, we obtain $e^{L/\xi}<10^6$
with the assumption of $t_{\mathrm{n.n.}}\sim 10^{15}\ {\rm s}^{-1}$.
This also assures a negligibly small transmittance 
$T\sim e^{-L/\xi}\sim 10^{-6}$, 
and thus the small leak current and low dissipation.

The above scenario can be experimentally verified 
by preparing thin films of solid solution 
such as Pb(Fe$_{0.5}$Nb$_{0.5}$)O$_3$ and Pb(Sc$_{0.5}$Nb$_{0.5}$)O$_3$, 
which contains a difference in the local electronic level. 
Note that the solid-solution system having B-site ions in equal ratio 
becomes ordinary ferroelectrics but not a classical relaxor 
if they are prepared with an adequate slow-anneal 
process~\cite{Galasso,AsanumaJJAP}. 
It can be regarded as the quantum analogue of the relaxor:
In relaxor ferroelectrics~\cite{Cross87,science_relaxor,Horiuchi00} 
and pinned charge density waves~\cite{GrunerRMP}, 
the mechanism is due to a classical mesoscopic cluster formation analogous to spin glass systems.
In addition, the choice of the parameters in the present model 
are realistic for experiments,
because the critical thickness for ferroelectricity of Pb-based
perovskite thin films is in the range of 1.2 $\sim$ 4 nm
\cite{TybellAPL,FongPRL} and the capacitor area of a practical
device (2M-bit FeRAM) is down to 0.4224 $\mu$m$^2$ \cite{IEDM}.

We briefly mention the relevanec of the present theory 
to the charge pumping through a {\it metallic dot}~\cite{Marcus99} 
weakly connected with two leads. 
Changing potential heights between the dot and the neighboring leads, 
an unquantized charge is pumped from one lead to the other~\cite{Brouwer98}.
Here, the main source of the pumping is a finite transmittance. 
It crosses over to the opposite regime with a vanishing transmittance 
where the pumped charge is nearly quantized 
around the resonant transmission~\cite{resonant}. 

In conclusion, we studied the quantum-mechanical disorder-enhanced
dielectric response in nanoscale insulators. 
Phase winding of the reflection coefficient $r$ at the vortex 
exists in the space of ionization and dimerization, and it plays 
a crucial role in adiabatic charge pumping 
and changing the polarization of the system through the resonance
tunneling, whose position varies with the chemical potential.
This serves as a new quantum theory of the FeRAM or capacitance.

The authors would like to thank S. Horiuchi, Y. Okimoto, Y.
Tokura, and D. Vanderbilt for discussions. The work
was partly supported by Grant-in-Aids 
(Grant No. 15104006, No. 16076205, and No. 17105002)
and NAREGI Nanoscience Project from the Ministry of 
Education, Culture, Sports, Science, and Technology.

\end{document}